\documentclass[fleqn,twoside]{article}
\usepackage{espcrc2}

\usepackage{graphicx}
\usepackage[figuresright]{rotating}
\usepackage{caption2}

\title{Ageing test of the ATLAS RPCs at X5-GIF}

\author{
        \centering
        \mbox{G. Aielli\address[Roma2]{INFN Sezione di Roma II, via della Ricerca Scientifica 1, 00133 Roma, Italy}},
        \mbox{M. Alviggi\address[Napoli]{INFN Sezione di Napoli; Complesso Univ. M.te S. Angelo, via Cinthia ed. G, 80126 Napoli, Italy}},
	\mbox{V. Ammosov\address[Pro]{IHEP, Protvino, Russia}},
        \mbox{M. Biglietti\addressmark[Napoli]},
        \mbox{P. Camarri\addressmark[Roma2]},
        \mbox{V. Canale\addressmark[Napoli]},
        \mbox{M. Caprio\addressmark[Napoli]}, 
        \mbox{R. Cardarelli\addressmark[Roma2]},
        \mbox{G. Carlino\addressmark[Napoli]},
	\mbox{G. Cataldi\address[Lecce]{INFN Sezione di Lecce; via Arnesano, 73100 Lecce, Italy}},
        \mbox{G. Chiodini\addressmark[Lecce]},
        \mbox{F. Conventi\addressmark[Napoli]},  
        \mbox{R. de Asmundis \addressmark[Napoli]},  
        \mbox{M. Della Pietra\addressmark[Napoli]},
        \mbox{D. Della Volpe\addressmark[Napoli]},
        \mbox{A. Di Ciaccio\addressmark[Roma2]},
        \mbox{A. Di Simone\addressmark[Roma2] \thanks{Corresponding author. Tel. +39 062023644. {\em E-mail}: Andrea.DiSimone@roma2.infn.it}},
        \mbox{L. Di Stante\addressmark[Roma2]},
        \mbox{E. Gorini\addressmark[Lecce]},
        \mbox{F. Grancagnolo\addressmark[Lecce]},  
        \mbox{P. Iengo\addressmark[Napoli]},
        \mbox{B. Liberti\addressmark[Roma2]},
	\mbox{A. Nisati\address[Roma1]{INFN Sezione di Roma I; P.zza Aldo Moro 5, Roma, Italy}},
	\mbox{Fr. Pastore\addressmark[Roma1]},
        \mbox{E. Pastori\addressmark[Roma2]},
        \mbox{S. Patricelli\addressmark[Napoli]},  
        \mbox{R. Perrino\addressmark[Lecce]},
        \mbox{M. Primavera\addressmark[Lecce]},     
        \mbox{R. Santonico\addressmark[Roma2]},
        \mbox{G. Sekhniaidze\addressmark[Napoli]},
        \mbox{S. Spagnolo\addressmark[Lecce]},
        \mbox{Yu. Sviridov\addressmark[Pro]},
	\mbox{R. Vari\addressmark[Roma1]},
	\mbox{S. Veneziano\addressmark[Roma1]},
        \mbox{V. Zaets\addressmark[Pro]}.
       }

\begin{document}

\begin{abstract}

An ageing test of three ATLAS production RPC stations is in course
at X5-GIF, the CERN irradiation facility.  
The chamber efficiencies are monitored using cosmic
rays triggered by a scintillator hodoscope.
Higher statistics measurements are made when the X5
muon beam is available. 
We report here the measurements of the efficiency versus operating voltage
at different source intensities,
up to a maximum counting rate of about 700Hz/cm$^2$.
We describe the performance of the chambers during the test
up to an overall ageing of 4 ATLAS equivalent years correspondong to an
integrtaed charge of 0.12C/cm$^2$, including a safety factor of 5.

\vspace{1pc}
\end{abstract}

\maketitle

\section{Experimental setup}

The Gamma Irradiation Facility, GIF, located downstream of the final dump
of the X5 beam, uses a $^{137}$Cs source of 20 $Ci$
to produce a large flux of 660 $keV$ $\gamma$ rays. 
A system of lead filters allow to reduce the flux up to a factor of $10^4$.
The X5 muon beam can also be sent into the area.
A more detailed description of this facility and of the
characteristics of the $\gamma$ flux can be found in
\cite{GIF}.
\newline

Three production ATLAS RPC chambers (BML-D) are installed in the area,
along the beam line.
The chambers have 2 detector layers which are read out by strips oriented
in both the $\eta$ and $\varphi$ directions
The chambers are perpendicular to the beam line, with the long side (about 4m)
oriented in the vertical direction.
For details on the chamber structure, see \cite{mutdr}.
The chambers are operated with the ATLAS gas mixture
$C_{2}H_{2}F_{4}/i \textendash C_{4}H_{10}/SF_{6}=94.7/5/0.3$.
\label{trigger}
\newline

The trigger is provided by the coincidence of three scintillator layers of
33x40 $cm^2$, each made of three slabs.
During the beam runs, the three layers are aligned along the beam line,
while for the cosmic rays runs, they
are arranged as a telescope with the axis oriented at 40 degrees with respect to
the vertical direction, in order to maximize the trigger rate.
\newline
Signals from the frontend electronics are sent to a standard ATLAS ``splitter board'' \cite{lvl1tdr} 
as in the final architecture foreseen for the trigger and readout electronics
, and subsequently to TDCs
working in common stop mode (see \cite{TDC}) which record up to 16 hits per
event per channel, in a 2$\mu s$ gate.
Both the leading and falling edges of the signals are recorded.
The data acquisition is performed using a LabView\texttrademark application.
\newline

The DCS system, also implemented using LabView\texttrademark, records both the
low and and high voltages as well as the gap currents.
Gas composition, together with all relevant environmental data such
as pressure, temperature and relative humidity are controlled as well.
The gas relative humidity, also monitored by the DCS, is set in the range of 30\%-50\% by bubbling in water
a fraction of the total gas flux.

\section{Ageing status}
The integrated ageing from the beginning of irradiation (November 2002)
up to July 2003, is equivalent to 4.1 ATLAS years.
The ageing is evaluated assuming a total charge per count of 30$pC$ and a counting rate
5 to 10 times larger than expected from the background simulations
in the ATLAS cavern.
Two main ageing parameters were measured during the test: the electrode
plate resistivity, which was shown in a previous test (\cite{mod0})
to gradually increase with the integrated detector current, and the noise
current, which is an indicator of possible degradations of the electrode surface.
\newline

\begin{figure}[t]
\includegraphics[scale=0.45,angle=-90]{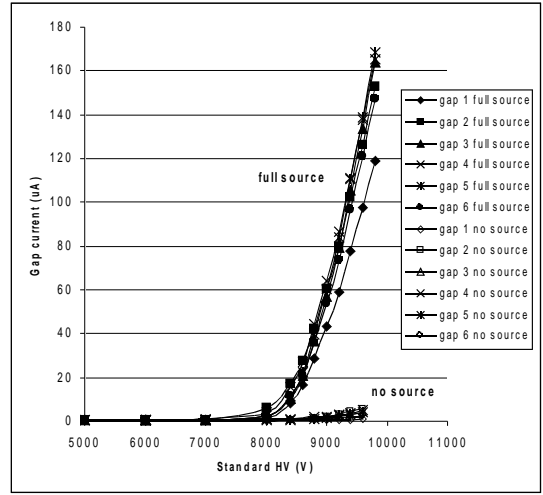}
\caption{Gap currents versus operating voltage at full irradiation and source off.}
\label{fig:correnti}
\end{figure}

\section{Experimental results}

All along the test, the chamber efficiencies are
monitored using cosmic rays.
The value of the detector plate resistivity are also periodically measured,
as discussed below. Gap currents are
measured on a daily basis, both with fully open source
and closed source. A typical plot of the current versus operating voltage is
shown in figure \ref{fig:correnti}.\newline
After 3.5 Atlas years (about 100 mC/cm$^2$ integrated charge)
accurate efficiency measurements
have been performed using the X5 muon beam.

The plate resistivity is a crucial parameter for RPC performance because
it determines the detector rate capability.
Two different methods were followed for the resitivity measurements:

\begin{figure}[t]
\includegraphics[scale=0.4]{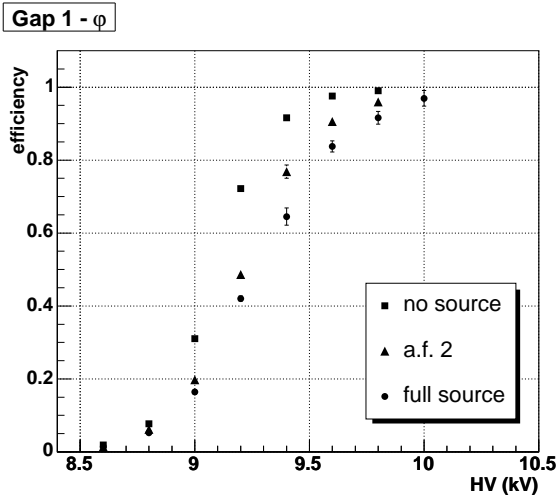}
\caption{Efficiency versus high voltage at different source intensities,
as measured with the muon X5 beam.}
\label{fig:eff}
\end{figure}

{\em HV drop correction.} This method is based on the comparison of the
efficiency plateaus with and without source irradiation. As shown in figure
\ref{fig:eff}, the current driven under irradiation is much higher
than the one just due to the muon beam or cosmic rays. This curent produces
a significant voltage drop across the plates and the 
voltage applied to the gas can be written (see \cite{Giulio}) as

\begin{equation}
V_{gas} =V - R_{pl}I_{gap}
\end{equation}

where $V_{gas}$ is the effective voltage on the gas gap, V is the power supply
voltage and I is the current driven by the gap.

This drop causes the efficiency plateaus under irradiation to be
shifted at higher HV values, as shown in figure \ref{fig:eff}. 
This shift, allows to evaluate the plate
resistivity when the gap current is also measured.
\newline

\begin{figure}[t]
\includegraphics[scale=0.40,angle=-90]{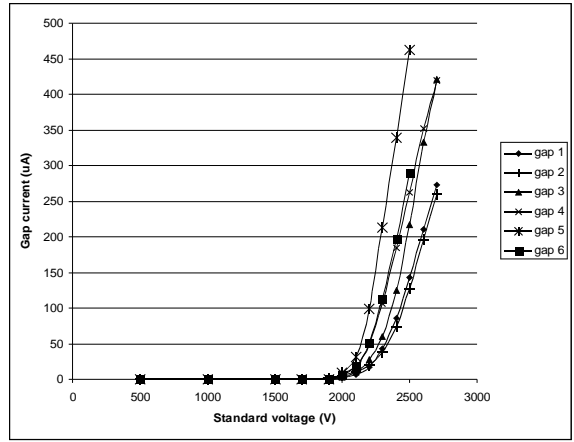}
\caption{I-V characteristics in pure Argon.}
\label{fig:argon}
\end{figure}

{\em I-V characterstic in pure Argon.} The standard ATLAS gas mixture is replaced
by pure Ar. As shown in figure \ref{fig:argon}, the I-V curve is
characterized in this case by a transition
region around V=2$kV$ with a fast current increase. For higher voltages
a linear current increase is observed.
We assume that above the transition
voltage, the drop across the Argon remains constant, so that the
slope $\Delta V/\Delta I$ 
in the linear region gives the plates total resistance.
\newline

During the test, the chamber efficiencies are monitored using cosmic rays
triggered by the scintillators.
High statistics efficiency measurements are made when the X5 muon beam is available.
\newline

For each event, the muon trajectory is reconstructed.

\ \newline
The scintillator slabs of the trigger system, of size
11x40$cm^2$, would only allow a very modest tracking, not sufficient to eliminate
all accidental hits at very high counting rate induced by the source.
The tracking capability is therefore improved using the hits recorded 
by the RPCs, as decribed below:\newline
(1) Hits recorded by the RPC under test are ignored in the tracking.\newline
(2) Only hits registered in a time window of 25ns around the beam
peack are considered useful for the track reconstruction.\newline
(3) A track is considered reliable only if it is built upon at least
three read out layers (out of five) with only one cluster. \newline
(4) A reliable tracking in both the $\eta$ and $\varphi$ directions
is required. \newline
(5) The layer is considered efficient if it shows a
cluster aligned to the track within $\pm$ 1 strip.
\newline

\begin{figure}[htb]
\includegraphics[scale=0.40]{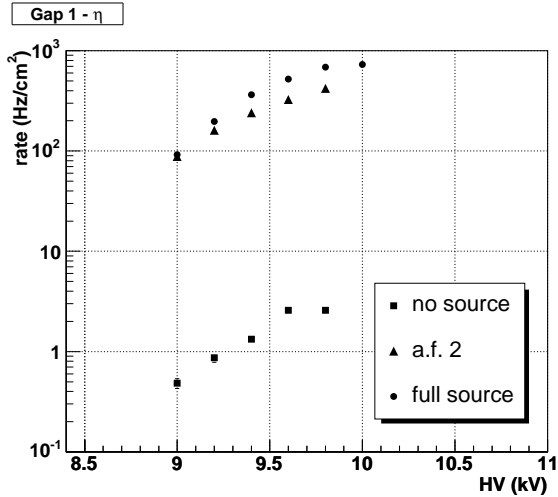}
\caption{Single rate measurement at different source intensities for gap 1.}
\label{fig:rate}
\end{figure}

High statistics measurements were made with the X5 beam.
Data sets were taken with different front end
electronic thresholds, source intensities, and operating voltages.
The plateaus shown in figure \ref{fig:eff}, for example, refer to the gap
number 1, which is the nearest to the source (at about 2.5$m$). \newline
Figure \ref{fig:rate} shows, for the same gap, the counting rate as a function
of the HV for different irradiation intensities.

\section{Gas recirculation}

At the beginning  of July 2003 the gas closed loop was introduced on 4 out of the
6 tested gas gaps, in order to simulate the real working conditions in ATLAS.
The two remaining gaps were left in open flow for comparison.
The output gas of the chambers goes through a system of filters
and then is sent again in the gaps, together with a fraction of fresh gas.
The total gas flow is set at \mbox{40 $l/h$}, and the fraction of recirculated
gas has been gradually increased as reported in table 1.
The gas at the input and at the output of the chambers was sampled and
analyzed in order to find any polluting component related to the chamber operation.
The analisys of the recirculated gas didn't show any excess of
pollutant with respect to open flow operation.

\begin{table}[htb]
\label{tab:gas}
\centering
\begin{tabular}{|r|l|}
\hline
time & fraction of\\
 & recirculated gas\\
\hline \hline
first 3 months & 50\% \\ 
following 3 days & 80\% \\ 
following 5 days & 90\% \\ 
presently & 95\% \\
\hline
\end{tabular}
\vspace{10pt}
\caption{Fraction of recirculated gas with respect to total flux}
\end{table}

\section{Conlusions}

The measurements performed at X5-GIF on three ATLAS production RPCs
have shown an increase of the electrode plate resistivity, as expected
from previous tests (\cite{mod0}).
The rate capability of all tested chambers remains, however, much above
the ATLAS requirements.
Indeed, all the gaps under test showed a very good detection efficiency even
at fully opened source, with a counting rate of about 700$Hz/cm^2$.
\newline
An increase of the source-off currents was observed during the
summer.
This effect, due to the extreme conditions of the test (about
40 times the nominal ATLAS rate), has been shown to be amplified by
low gas flow and operating temperature above 30$^{\circ}C$.
This effect was shown to be to some extent reversible, when the gas
flow and temperature conditions are set again to acceptable values
and the ageing rate is reduced.
Ohmic currents didn't show any significant increase all along the
ageing test, except the variations due to temperature excursion.

\end{document}